# Numerical Study of Non-Newtonian Effects on Fast Transient Flows in Helical Pipes

Mohsen Azhdari

Department of Mechanical Engineering, University of Tehran, Tehran 11155-4563, Iran

Email: m.azhdari@ut.ac.ir

Alireza Riasi

Department of Mechanical Engineering, University of Tehran, Tehran 11155-4563, Iran

Email: ariasi@ut.ac.ir

Pedram Tazraei

Department of Mechanical Engineering, Texas A&M University, College Station, TX 77843-3120, USA

Email: ptazraei@tamu.edu

**Abstract**

*This study focuses on a parametric study of the laminar fast transient flow of non-Newtonian fluids through helical pipes. Classical simulations of fluid hammer do not deal with the pipeline helicity and non-Newtonian characteristics of the fluid, while the present work addresses those features. To this end, the power-law model is employed to accommodate the non-Newtonian behavior of the fluid. Effects of the pipe wall elasticity and compressibility of the working fluid are taken into account through a modified bulk modulus elasticity of the fluid. The results of the three-dimensional numerical analysis followed herein demonstrate good agreement with the available experimental data, and they show that non-Newtonian properties of the fluid significantly influence the pressure head response, velocity and shear stress profiles, and also the strength of the formed secondary flows. At the first stage of the fluid hammer, where the maximum deviation arises, the magnitude of the wall shear stress at the pipe midpoint for the shear-thinning and shear thickening fluids are respectively 67.7% lower and 200% higher than the Newtonian fluid. Furthermore, the average magnitude of the axial vorticity over the upper half of the pipe cross-section area for the shear-thinning and shear-thickening fluids are respectively 65.5% lower and 111.7% upper than the Newtonian case.*

## 1. INTRODUCTION

In pressurized conduits, changes in boundary conditions cause some perturbations in the fluid to be generated in order to regulate the flow rate. These changes may occur due to different factors including start-up or failure of the pump and turbine or fast opening or closing of the valve, to name a few. As a result, a series of positive and negative pressure waves propagate along the passage with the wave speed which, in turn, cause serious problems such as pipe failure, hydraulic equipment damage, corrosion, and cavitation. This phenomenon is then referred to as "water-hammer," if the working fluid is water, otherwise, is called "fluid-hammer."

Previous studies have mostly focused primarily on the transient flow of Newtonian fluids through straight pipes, and have not investigated how helicity and non-Newtonian features influence flow behavior. Ghidaoui *et al.* [1] presented a literature review on historic developments and novel research and practices in the field of hydraulic transients. The behavior of unsteady velocity profiles in transient flow problems has already been investigated by several researchers such as Silva-Araya and Chaudhry [2], Ghidaoui and Kolyshkin [3], Wahba [4], and Riasi *et al.* [5]. However, the non-Newtonian fast transient flows in straight pipes have received, comparatively, much less attention. For instance, Wahba [6] used the power-law model to investigate shear-thinning and shear-thickening effects of fluid on the transient flow behavior. Tazraei and Riasi [7] also introduced a desirable approach using the Carreau model and reported how non-Newtonian effects could significantly influence the flow behavior through straight pipes.

Secondary flows which are induced as a result of fluid flowing in curved tubes have significant ability to enhance the heat transfer rate due to the mixing of the flow. Furthermore, the compact structure of helical pipes allows them to be used in many applications such as chemical reactors, power generators, heat exchangers and refrigerators. Dean [8,9] was the first who mathematically analyzed the problem of flow in a coiled pipe, in which the steady flow of an incompressible fluid through a coiled pipe was considered. It was found out that the flow characteristics depend upon a single parameter, the Dean number which is defined as $Dn = Re\sqrt{\lambda}$, where $Re$ is the Reynolds number and $\lambda$ is the normalized curvature ratio that equals to $d/D$; $d$ and $D$ are the diameter of the pipe and coil, respectively. An extensive review of the fluid flow in curved tubes has been done by Berger *et al.* [10]. Recently, a great deal of attention has been paid to secondary flows formed in helical pipe flows. Germano [11] found that helicity of the coils greatly influences the secondary flow. It was also proposed that this effect is characterized by the dimensionless Germano number, $Gn = \eta Re$, in which, $\eta$ is the normalized torsion that equals to $\eta = h/\pi D$, and $h$ is the coil pitch. Liu and Shijie [12] pointed out that in addition to the Dean and Germano numbers, a third dimensionless group has an effect on the secondary flow pattern in such a way that for $Dn \geq 20$ it is defined by $\gamma = \eta/(\lambda Dn)^{\frac{1}{2}}$, and for $Dn < 20$, $\gamma = \eta/(\lambda Re)$.

Despite the importance of non-Newtonian flow fields through helical pipes in the polymer industry, biomedical science, and biochemical processing, relatively few such studies have been reported to date. Tazraei *et al*. [13] studied the blood-hammer phenomenon through the posterior cerebral artery and discussed how the velocity and wall shear stress profiles are affected by the non-Newtonian properties of blood. Some other works have been devoted to the study of shear-thinning blood flow with an emphasis on the curved path of arteries. The importance of these studies is that arterial flows, in reality, experience substantial secondary flows originating from the curvature of the flow path, varying cross-section, bifurcations, or surface protrusions. So their results give a general idea of the effects that one may come across in real arteries. The flow of a power-law fluid through a curved pipe has been analytically studied by Agrawal *et al.* [14]. They used the perturbation method to find an approximate analytical solution of the axial velocity and the stream function. Gijsen *et al.* [15] examined the non-Newtonian unsteady flow in a curved tube both numerically and experimentally. They employed the Carreau-Yasuda model to incorporate shear-thinning behavior of the blood and found a significant discrepancy in velocity profiles between the Newtonian and non-Newtonian cases. Cherry and Eaton [16] simulated the blood flows with Newtonian and non-Newtonian models in straight and curved paths. They modeled shear-thinning viscosity as a function, which fits the experimental data, of the local shear rate and compared the strength of secondary flows between the Newtonian and non-Newtonian cases.

Since the literature lacks a comprehensive study of the non-Newtonian fast transient flows through helical pipes, the main purpose of the current study is to bridge the gap between the transient flows in helical pipes and the fluid-hammer theory. This is done through employing the power-law model to describe the viscosity of the non-Newtonian fluid, and using a three-dimensional numerical model to solve the equations governing the problem. In this way, a parametric study has been conducted in order to investigate the non-Newtonian effects of the fluid on the behavior of unsteady velocity and shear stress profiles, the pattern of the Dean vortex, and the strength of the secondary flows.

## 2. MATHEMATICAL MODELING AND NUMERICAL PROCEDURE

Governing equations describing unsteady compressible flows in full three-dimensional form can be expressed as the following continuity and momentum equations. Using Einstein's tensor notation, the governing equations read as follows [17]

$$\frac{\partial \rho}{\partial t} + \frac{\partial}{\partial x_i}(\rho u_i) = 0 \qquad (1)$$

$$\frac{\partial}{\partial t}(\rho u_i) + \frac{\partial}{\partial x_j}(\rho u_i u_j) = \frac{\partial}{\partial x_j}(\tau_{ij}) \qquad (2)$$

where $i$ and $j$ are indices running from 1 to 3 and represent the three spatial directions, $x$ is the spatial coordinate, $\rho$ is the fluid density, $t$ is the time, $u$ is the velocity component and $\tau$ is the shear stress. The acoustic wave speed considering the effects of both fluid compressibility and pipe flexibility is defined as

$$a^2 = \frac{K/\rho}{1 + (K/E)(D/e)} \qquad (3)$$

Where $E$ is the Young's modulus of the pipe wall, $K$ is the bulk modulus of the fluid, $D$ is the pipe diameter, and $e$ is the pipe wall thickness. The density can be expressed in terms of the created overpressure in the flow as

$$\rho = \frac{\rho_0}{K}\Delta P + \rho_0 \qquad (4)$$

in which, $\rho_0$ is the density of the fluid at the undisturbed state, and $P$ is the static pressure. For the power-law fluid, the general form of the constitutive equation can be expressed as [18]

$$\tau_{ij} = m\left|II_{2D}\right|^{(n-1)/2}(2D_{ij}) \qquad (5)$$

wherein, $n$ is the power-law index. For $n<1$ the fluid is shear thinning, whereas for $n>1$ the fluid is shear-thickening. For the special case of Newtonian fluid ($n=1$), the consistency index $m$ is identically equal to the dynamic viscosity of the fluid, $\mu$. $2D_{ij}$ is the rate of deformation tensor defined as Eq. 6, and $\left|II_{2D_{ij}}\right|$ is the second invariant of $2D_{ij}$.

$$2D_{ij} = \left[\frac{\partial u_i}{\partial x_j} + \frac{\partial u_j}{\partial x_i}\right] \qquad (6)$$

At the steady state, the generalized Reynolds number is given by [19]

$$Re' = \frac{8\rho U_0^{2-n} D^n}{m(6+2/n)^n} \qquad (7)$$

for which, $U_0$ is the average of axial velocity over the pipe cross-section. In this paper, flow through a helical pipe, which is made up of copper is investigated and the friction factor can be defined in terms of the generalized Reynolds number,

$$f_0 = \frac{2\tau_{w,0}}{\rho U_0^2} = \frac{16}{Re'} \qquad (8)$$

A commercial CFD code has been employed, through which the discretized equations are solved in a segregated manner. In addition, the Semi-Implicit Method for Pressure-Linked Equations (SIMPLE) algorithm which is one of the most common pressure-velocity coupling algorithms is used herein to compute the pressure from the momentum and continuity equations. Also, the second order upwind spatial discretization and the common first-order implicit time integration scheme are employed to integrate the mass and momentum equations. The details of the case studies are listed in Table 1. According to Fig.1, showing the schematic and boundary conditions of the coiled pipe, $O$ is the center of the helical pipe at a certain cross-section. $A$ and $B$ represent, respectively, inner and outer sides of the helical pipe. ($A$-$B$) plane represents the pipe midpoint which is divided into ($A$-$O$) and ($O$-$B$) portions corresponding to the inner and outer halves of the pipe cross-section, respectively. In the results section, velocity and shear stress profiles are plotted at the pipe midpoint-($A$-$B$) plane- for different power-law indices. The consistency index is assumed to be equal to the dynamic viscosity of water, $m = \mu = 0.001\,\text{Pa.s}$. Generalized Reynolds number for all cases lies in the laminar regime, $\text{Re}' = 670$. As shown in Fig. 1, to specify appropriate boundary conditions, pressure at the upstream end of the pipe is taken equal to the upstream reservoir pressure head, $H = 32\,\text{bar}$. At the closed downstream valve, the outlet pressure at the steady state is predefined such that the mass flux within the pipe remains constant for all the cases tested, and to imitate the sudden closure of the valve on the threshold of the fluid hammer, the axial velocity component is set to zero.

## 3. MODEL VERIFICATION AND VALIDATION

In this section, verification (numerical uncertainty) and validation (comparison with the experimental data) of the numerical scheme are carried out to ascertain its effectiveness.

### 3.1. Verification

The objective of verification is to check if the numerical accuracy is independent of the physical modeling [20]. Hence, according to Table 1, the case study for the unit power-law index has been considered. Three different mesh sizes and accordingly time step sizes including a course, fine, and finer grids have been tabulated in Table 2. Based on the grid convergence index (GCI), the uncertainty estimation detailed in Table 3 demonstrates that the numerical uncertainties are in the acceptable range. Similar GCI tests were performed on the shear-thinning and shear-thickening cases, which confirmed the grid and time-step independencies. Pressure and wall shear stress are the integral physical parameters in transient flow problems. Therefore, $\Phi_1, \Phi_2$, and $\Phi_3$ represent the values of these parameters respectively for finer, fine, and coarse meshes at the middle of the pipe. Moreover, due to the occurrence of the maximum deviation at $t = 3.6(L/a)$, the corresponding values at this time is considered in Table 3. $p$ is the observed order of method, $e_{21}$ is the dimensionless error, GCI denotes the grid convergence index, $u_{num}$ is the standard uncertainty in the simulation solution, and $\varepsilon_{21} = \Phi_2 - \Phi_1, \varepsilon_{32} = \Phi_3 - \Phi_2$.

### 3.2. Validation

A geometry exactly the same as the one used by Holmboe and Rouleau [21] was regenerated on which the approach, followed in this work, was implemented for the sake of validation of the model. Also, all the grid resolutions and other contributing factors applied herein have been verified according to the past subsection. Fig. 2(a) and 2(b) compare the calculated pressure response with the experimental data reported in [21] at the valve and pipe midpoint, respectively. In this Figure, the time axis is normalized by the wave travel time $(L/a)$, and the pressure axis is normalized by the Joukowsky head $(AU_0/g)$. As is evident, the numerical results are in good agreement with the measured data and the simulation accurately resolves both the pressure amplitude and pressure phase.

### 4. NUMERICAL RESULTS

This section sheds light on how variations of the power-law index influence the pressure response, velocity and shear stress profiles, and secondary flows in the case of fluid-hammer.

### 4.1. Pressure Head Response

Fig. 3(a) and 3(b) show the pressure head response for different values of the power-law index at the downstream valve and pipe midpoint, respectively. According to Fig. 3, pressure response of the shear-thickening fluid experiences the highest decay rate compared to the Newtonian and shear-thinning fluids. For instance, at $t = 13(L/a)$, the value of pressure head at the valve point and pipe midpoint corresponding to the shear-thinning fluid is approximately 10% higher than that of the Newtonian fluid. Conversely, at the same time, the pressure head response for the shear-thickening fluid is less than the Newtonian case by 25%.

### 4.2. Velocity Profiles

The velocity profiles corresponding to the Newtonian case during the first wave cycle at the pipe midpoint are shown in Fig. 4. Figs. 4(a) and 4(b) pertain to the outer half (*O-B* plane) and the inner half (*O-A* plane) of the cross-section, respectively. The velocity axis is normalized by the initial bulk velocity ($U_0$) and the position axis is normalized by the pipe radius (*R*). Negative values on the position axis represent the inner half (*O-A* plane), and positive values denote the outer half (O-B plane) of the pipe cross-section. When it comes to the velocity profile, asymmetry of the profiles is the main difference between the flow through the straight and helical pipes in that the velocity profiles are asymmetric in the helical pipes. Figs. 5 and 6 show the velocity profiles at the pipe midpoint at $t = 0.6(L/a)$ and $t = 3.6(L/a)$, respectively. These two figures depict the influence of the power-law index variation on the velocity profile. As shown, as the power-law index decreases, the location at which the maximum axial velocity happens shifts toward the pipe wall. This, in turn, implies that the region of reverse flows and strong gradients gets closer to the pipe wall.

### 4.3. Shear Stress

Shear stress profile corresponding to the Newtonian fluid during the first wave cycle at the pipe midpoint is shown in Fig. 7. It is worth noting that the shear stress is normalized by $(\rho_0 U_0 aR/L)$. When the positive pressure wave travels from the valve to the reservoir and is reflected from the reservoir to the valve $(0 < t < 2L/a)$, shear stress near the wall is going to be in the opposite direction from the steady state shear stress. In this way, the fluid near the wall experiences negative values of the shear stress. But, when the working fluid is affected by the negative pressure wave traveling from the valve to the reservoir and also its reflection from the reservoir to the valve $(2L/a < t < 4L/a)$, shear stress near the wall has the same sign as the steady state flow.

In Fig. 8, the dimensionless wall shear stress of the Newtonian case versus the angular position at the pipe midpoint is shown. It also demonstrates that the absolute magnitude of the wall shear stress in transient flow is significantly higher than that of the steady flow irrespective of the fluid index. To quantitatively evaluate the non-Newtonian effects, contours of the dimensionless wall shear stress versus angular position in terms of the fluid indices at the pipe midpoint and at $t = 0.6(L/a)$ and $t = 3.6(L/a)$ are plotted in Fig. 9. According to Fig. 9, as the power-law index increases, the magnitude of the wall shear stress goes up. Also, regardless of the flow type, $\theta = 180°$ remains the symmetry axis of the shear stress contours. In addition, at $t = 0.6(L/a)$, the angle at which the minimum magnitude of the wall shear stress occurs is $\theta = 0°$. Absolute value of the shear stress uniformly increases to reach a maximum at $\theta = 180°$. On the contrary, this is not the case with the shear stress contour formed at $t = 3.6(L/a)$. At $t = 0.6(L/a)$, the maximum shear stress deviation occurs at $\theta = 180°$, i.e., the magnitude of shear stress corresponding to the shear-thinning and shear-thickening fluids are respectively 67.7% lower and 200% higher than the Newtonian fluid.

### 4.4. Secondary Flows

The curved shape of the helical pipes gives rise to the centrifugal force acting on the flow. Two factors determine the magnitude of the centrifugal force. The first one is the local axial velocity of the fluid particles and the second factor is the radius of curvature of the coil. In the near pipe wall, the fluid particles experience a lower centrifugal force than those which are in the core region of the cross-section. Thus, the fluid particles are pushed from the core region (*O*) toward the outer wall (*B*), where bifurcation occurs, and then appeared to be pushed toward the inner wall (*A*) along the pipe periphery, causing generation of the secondary flows composed of counter-rotating vortices known as Dean vortices.

The strength and pattern of secondary flows depend on several parameters. In the steady state flow through coiled pipes, the curvature of the path causes the Dean vortices to appear at the upper left and the lower right of the cross-section. After the fluid-hammer occurs, $0 < t < T/2$, change of the fluid velocity from $+U_0$ to $-U_0$ makes the centers of vortices smoothly rotate in the clockwise direction across the cross-section and is eventually placed at the upper right and lower left of the cross-section. Also, within the interval $T/2 < t < T$, due to the velocity shift from $-U_0$ to $+U_0$, centers of the Dean vortices move in the opposite direction until they are placed at the upper left and the lower right again, similar to the steady state.

Finally, the effect of non-Newtonian properties of the fluid on the shape of secondary flows pattern at the pipe midpoint can be seen in Fig. 10, in which the results are plotted at two different times, i.e., $t = 0.6(L/a)$ and $t = 2.6(L/a)$. According to Fig. 10, at a certain time, significant differences exist between the Newtonian and non-Newtonian fluids as to the shape of the secondary flows. As the power-law index decreases, the center of vortices shifts towards the pipe wall and also the strength of secondary flows decreases. The vortices formed at the upper and lower parts of the pipe cross-section are equal in strength but opposite in direction. To scrutinize the effects of non-Newtonian properties of the fluid on the strength of the secondary flows, Table 4 lists the magnitude of the axial vorticity averaged over the upper half of the cross-section at the pipe midpoint. According to which, at $t = 0.6(L/a)$, the magnitude of axial vorticity corresponding to the shear-thinning and shear-thickening fluids are respectively 65.5% lower and 111.7% higher than the Newtonian fluid.

## 5. CONCLUSIONS

A numerical study of non-Newtonian laminar transient flows resulting from the sudden closure of the downstream valve in a helical pipe has been reported, in which a commercial CFD code was used to solve the governing equations by a three-dimensional modeling. Non-Newtonian characteristics of fluids have been considered through employing the power-law model. Verification and validation of the solution were carried out in detail. Some of the main characteristics of the unsteady flow such as pressure head response, velocity and shear stress profiles, wall shear stress and secondary flows have been studied to compare the behavior of the flow with both the Newtonian and non-Newtonian working fluids when fluid-hammer occurs.

Based on the results, it can be reasonably concluded that the lower the fluid index, the lower the decay rate of the pressure head response. And the location corresponding to the maximum axial velocity gets close to the outer wall as the fluid index decreases. In addition, the shear-thinning fluid flows tend to have less secondary flow strength and wall shear stress magnitude compared to the Newtonian and shear-thickening ones.

**NOMENCLATURE**

| | |
|---|---|
| $a$ | acoustic wave velocity |
| $A$ | inner side of the pipe |
| $B$ | outer side of the pipe |
| $d$ | pipe diameter |
| $D$ | coil diameter |
| Dn | Dean number |
| $e$ | pipe thickness |
| $E$ | Young's modulus of elasticity of pipe wall |
| $f$ | friction factor |
| $g$ | gravity acceleration |
| Gn | Germano number |
| $h$ | coil pitch |
| $H$ | piezometric head |
| $k$ | bulk modulus of elasticity of fluid |
| $L$ | pipe length |
| $m$ | consistency parameter for the power-law model |
| $n$ | flow index for the power-law model |
| $O$ | center of the pipe |
| $P$ | static pressure |
| $R$ | pipe radius |
| Re | Reynolds number |
| Re′ | generalized Reynolds number |
| $t$ | Time |
| $u$ | velocity component |
| $U_0$ | average cross-sectional axial velocity at steady state |

| $x$ | spatial coordinate |
|---|---|

**Greek Symbols**

| $\gamma$ | dimensionless number |
|---|---|
| $\eta$ | normalized torsion |
| $\lambda$ | normalized curvature ratio |
| $\mu$ | apparent viscosity |
| $\rho$ | fluid density |

**Subscripts**

| $i$ | spatial direction |
|---|---|
| $j$ | spatial direction |
| $w$ | Wall |
| $0$ | steady state |

**Figure Caption List**

Fig. 1. Schematic and boundary conditions of the modelled coiled pipe

Fig. 2. Pressure-time history validation (a) at the valve, and (b) at the pipe midpoint

Fig. 3. Pressure head response versus time (a) at the valve, and (b) at the pipe midpoint

Fig. 4. Velocity profiles at the pipe midpoint for n=1 on the (a) *OB* plane, and (b) *OA* plane

Fig. 5. Velocity profiles at t=0.6L/a on the (a) *OB* plane, and (b) *OA* plane

Fig. 6. Velocity profiles at t=3.6L/a on the (a) *OB* plane, and (b) *OA* plane

Fig. 7. Shear stress profiles at the pipe midpoint for n=1 on the (a) *OB* plane, and (b) *OA* plane

Fig. 8. Variation of the wall shear stress versus angular position for n=1

Fig. 9. Variation of the wall shear stress versus angular positionfor (a) t=0.6 L/a and n=0.6, (b) t=0.6L/a and n=1, (c) t=0.6L/a and n=1.4, (d) t=3.6L/a and n=0.6, (e) t=3.6L/a and n=1, (f) t=3.6L/a and n=1.4

Fig. 10. Secondary flows at the pipe midpoint for (a) t=0.6 L/a and n=0.6, (b) t=0.6L/a and n=1, (c) t=0.6L/a and n=1.4, (d) t=2.6L/a and n=0.6, (e) t=2.6L/a and n=1, (f) t=2.6L/a and n=1.4

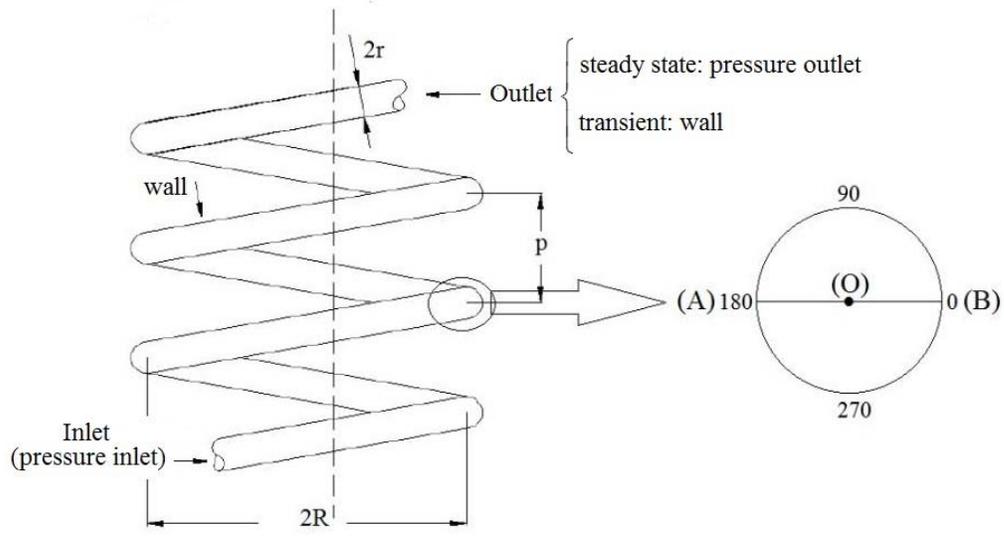

Fig. 1

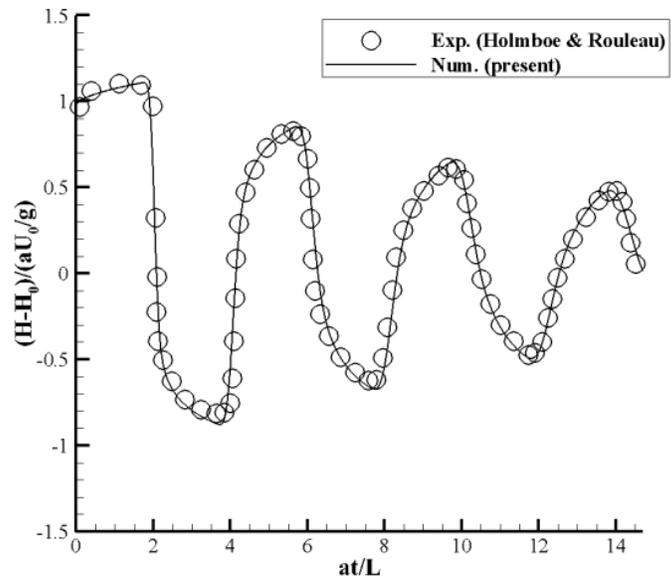

(a)

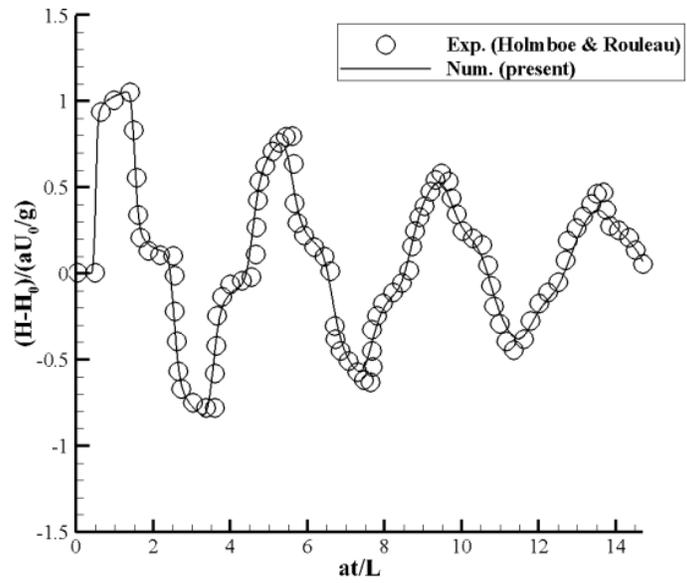

(b)

Fig. 2

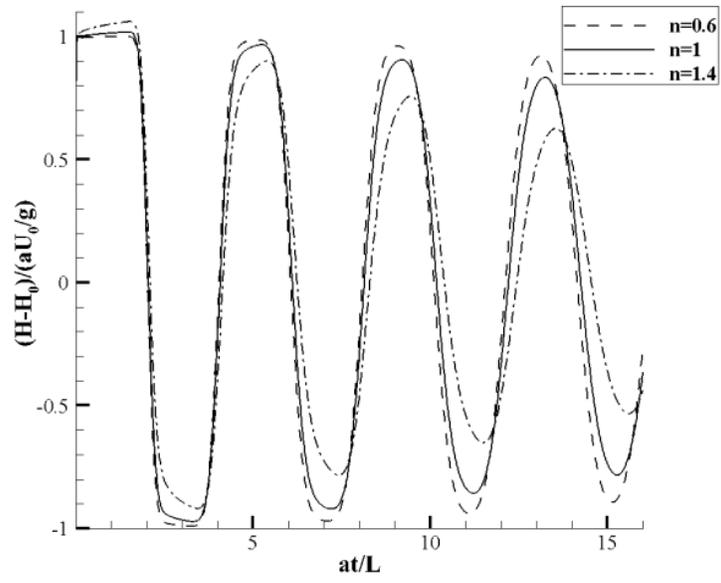

(a)

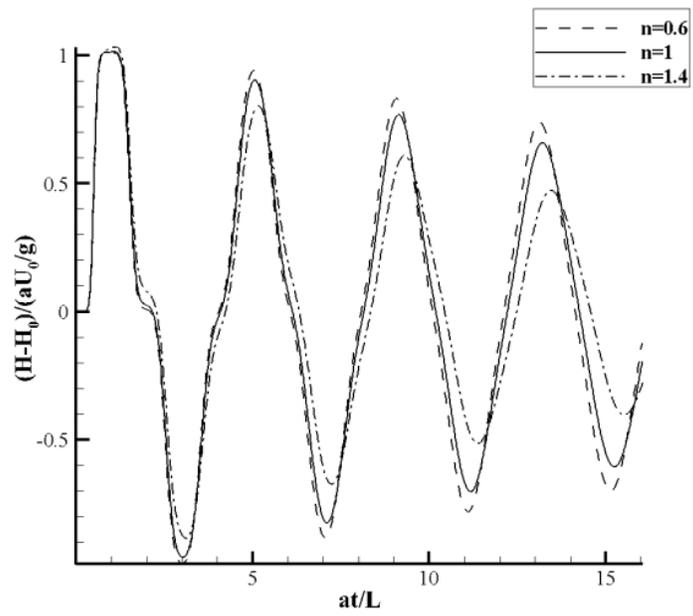

(b)

Fig. 3

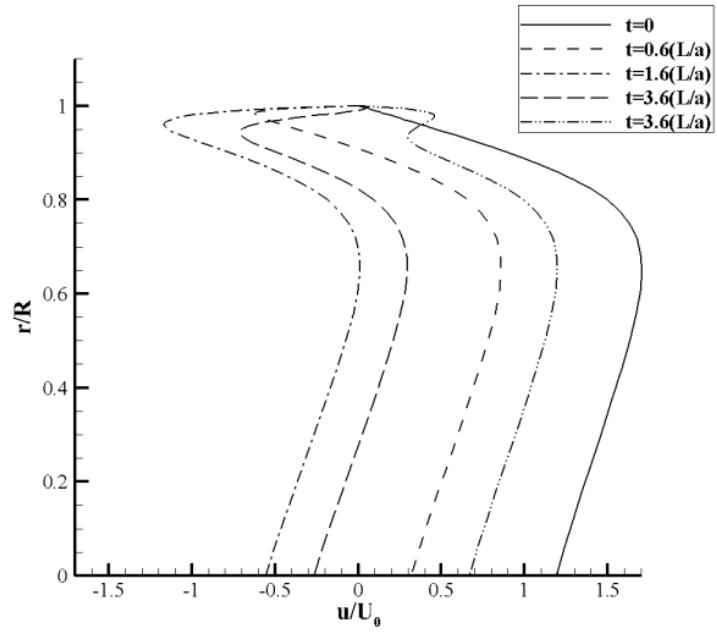

(a)

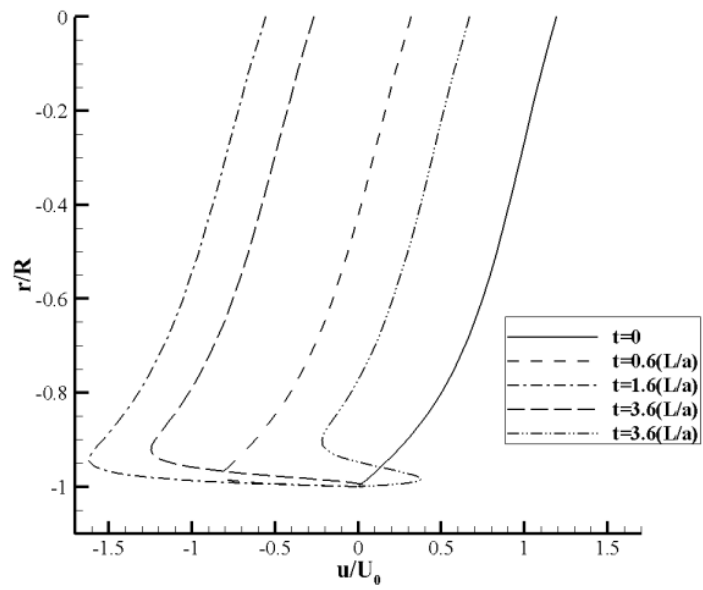

(b)

Fig. 4

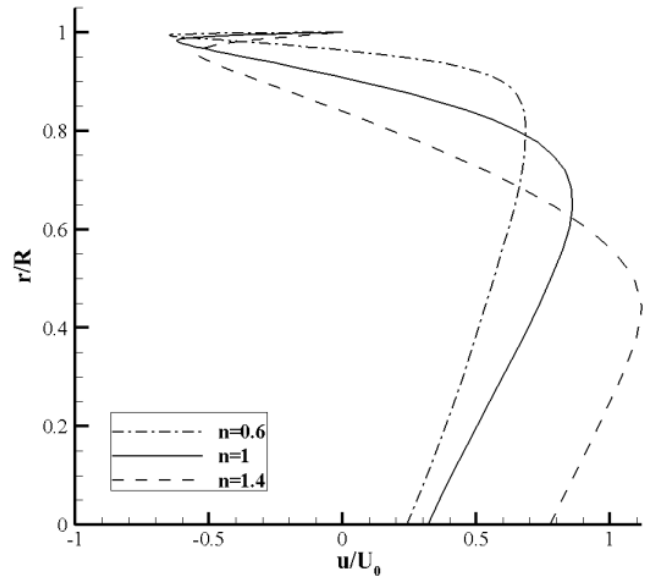

(a)

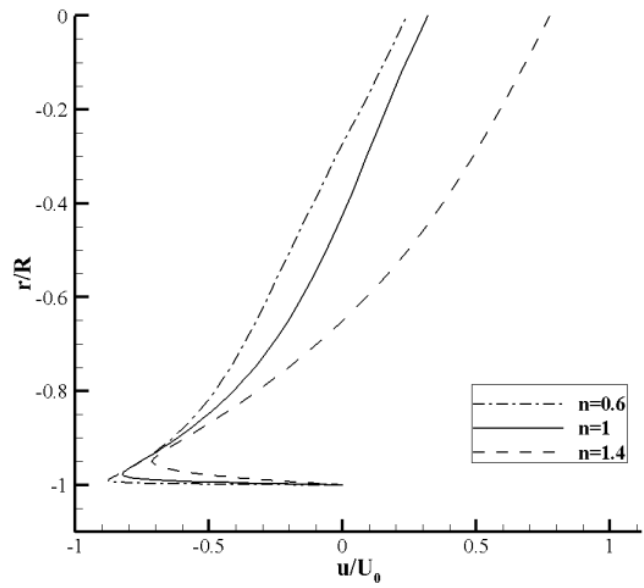

(b)

Fig. 5

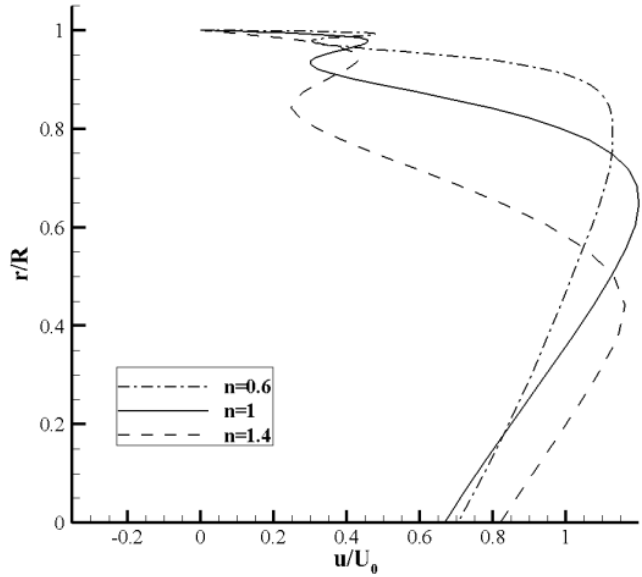

(a)

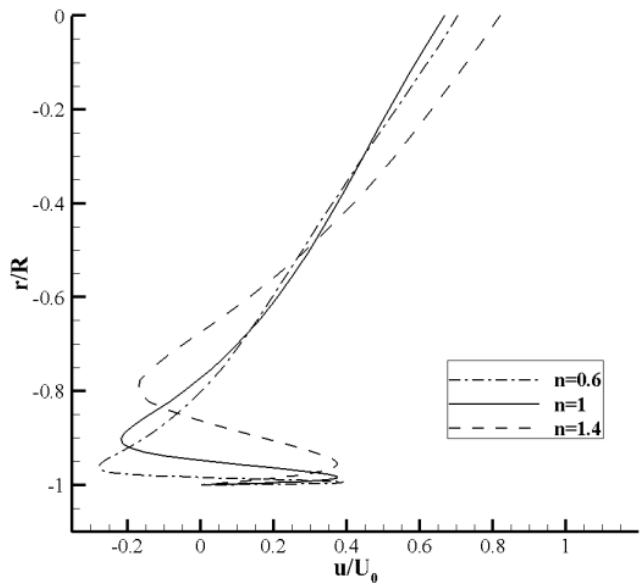

(b)

Fig. 6

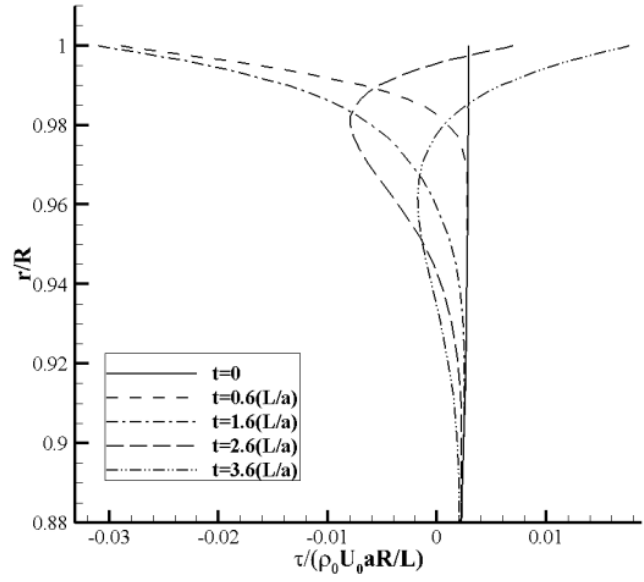

(a)

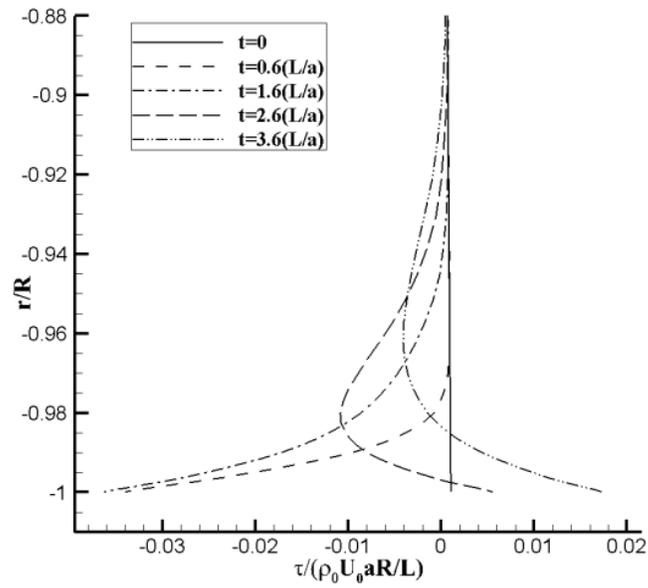

(b)

Fig. 7

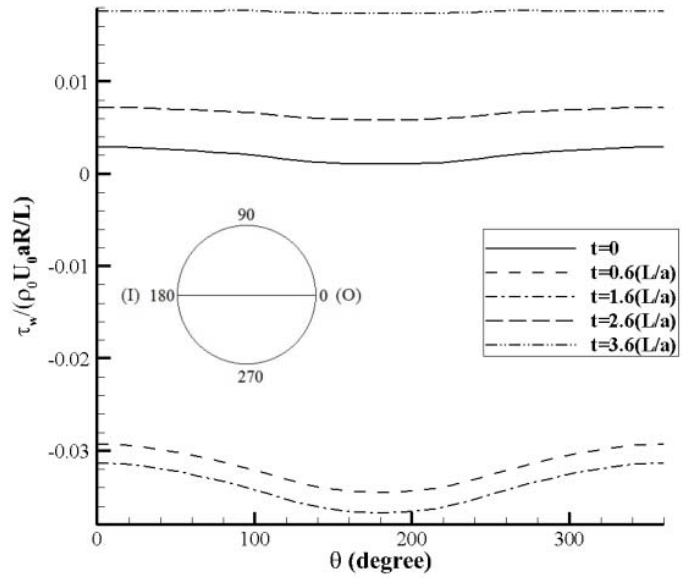

Fig. 8

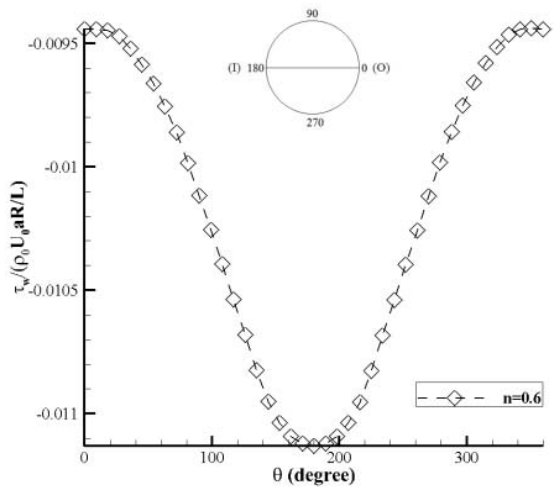

(a)

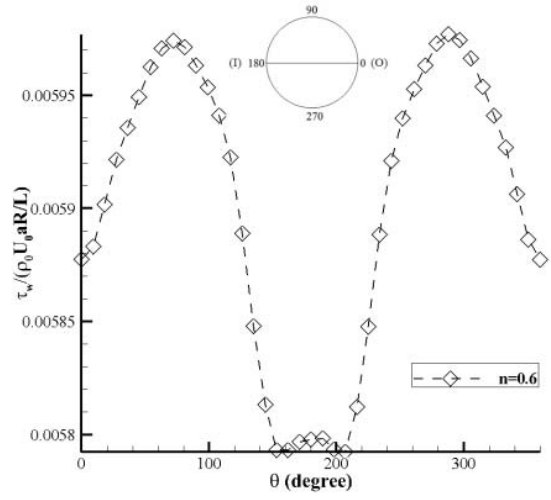

(d)

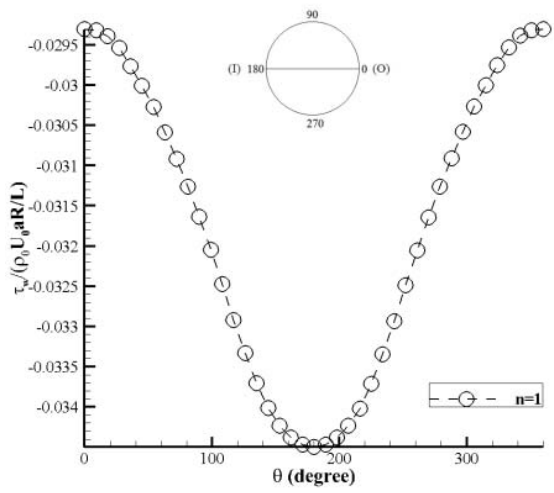

(b)

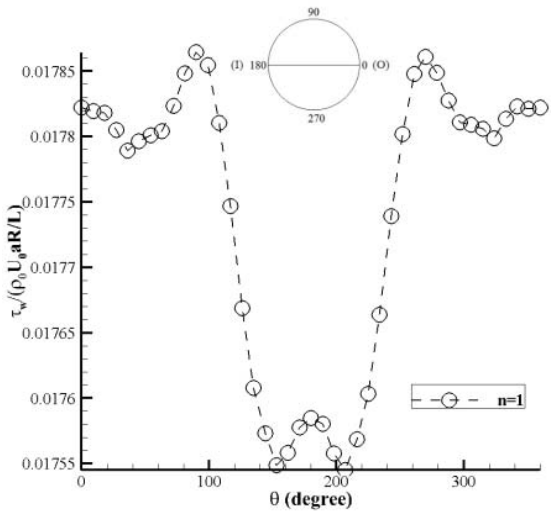

(e)

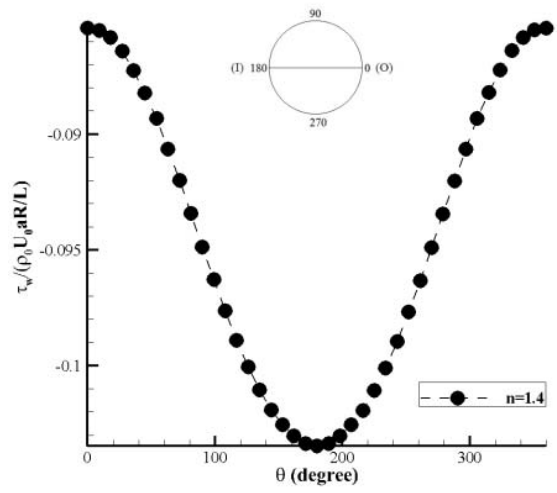 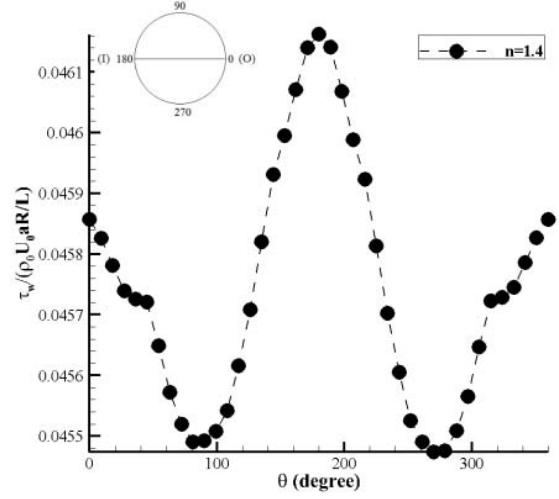

(c)                          (f)

Fig. 9

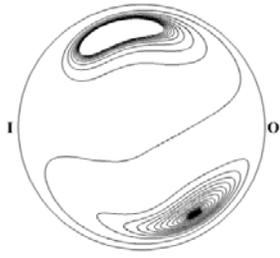 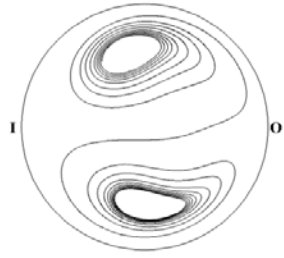 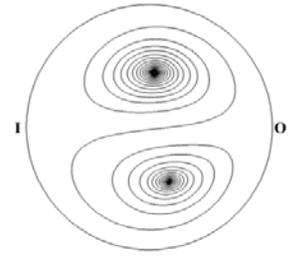

(a)       (b)       (c)

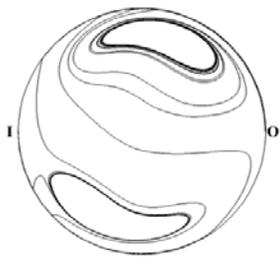 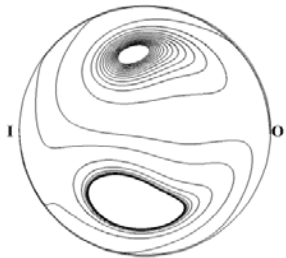 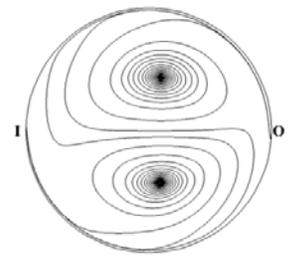

(d)       (e)       (f)

Fig. 10

**Table Caption List**

Table 1. The details of the test cases examined

Table 2. Grid independency study

Table 3. The details of the uncertainty estimation

Table 4. Strength of the axial vorticities formed in the upper half of the cross-section

Table 1

| Length of pipe (m) | Radius of pipe (m) | Radius of coil (m) | Pitch of coil (m) | Density $(kg/m^3)$ | Wave speed (m/s) | SS Dean number (–) | Generalized Reynolds number (–) | Consistency index $(Pa.s^n)$ | Power-law index (–) |
|---|---|---|---|---|---|---|---|---|---|
| 10 | 0.005 | 0.1 | 0.012 | 998.2 | 1329 | 149 | 670 | 0.001 | 0.6 |
| 10 | 0.005 | 0.1 | 0.012 | 998.2 | 1329 | 149 | 670 | 0.001 | 1 |
| 10 | 0.005 | 0.1 | 0.012 | 998.2 | 1329 | 149 | 670 | 0.001 | 1.4 |

Table 2

| Grid No. | Mesh Grid | Time Step (s) |
|----------|-----------|---------------|
| 1 | 461,158 | $1.3103 \times 10^{-4}$ |
| 2 | 1,049,600 | $1 \times 10^{-4}$ |
| 3 | 2,360,925 | $7.6767 \times 10^{-5}$ |

Table 3

|  | Joukowsky head (-) | Wall shear stress (Pa) |
|---|---|---|
| $\Phi_1$ | -0.4108 | 0.7900 |
| $\Phi_2$ | -0.4101 | 0.7745 |
| $\Phi_3$ | -0.4091 | 0.7536 |
| $\varepsilon_{21}$ | 0.0007 | -0.0155 |
| $\varepsilon_{32}$ | 0.001 | -0.0209 |
| $p$ | 1.3208 | 1.107 |
| $e_{21}$ | 0.17% | 1.96% |
| GCI | 0.497% | 7.03% |
| $u_{num}$ | 0.248% | 3.51% |

Table 4

| power index | $\omega_z(1/s)$ $t=0.6(L/a)$ | $\omega_z(1/s)$ $t=2.6(L/a)$ |
|---|---|---|
| n=0.6 | 0.2258 | 0.2117 |
| n=1 | 0.6465 | 0.6100 |
| n=1.4 | 1.3616 | 1.2107 |